# Imaging through multimode fibres with physical prior

CHUNCHENG ZHANG,[1,†] YINGJIE SHI,[1,†] XIUBAO SUI,[1,*] QIAN CHEN,[1]

[1] Nanjing University of Science and Technology, Nanjing 210094, China
[†]These authors contributed equally to this letter
*Corresponding author: sxb@njust.edu.cn

**Imaging through perturbed multimode fibres based on deep learning has been widely researched. However, existing methods mainly use target-speckle pairs in different configurations. It is challenging to reconstruct targets without trained networks. In this paper, we propose a physics-assisted, unsupervised, learning-based fibre imaging scheme. The role of the physical prior is to simplify the mapping relationship between the speckle pattern and the target image, thereby reducing the computational complexity. The unsupervised network learns target features according to the optimized direction provided by the physical prior. Therefore, the reconstruction process of the online learning only requires a few speckle patterns and unpaired targets. The proposed scheme also increases the generalization ability of the learning-based method in perturbed multimode fibres. Our scheme has the potential to extend the application of multimode fibre imaging.**

Multimode fibre (MMF) has been prominently used in many fields, such as laser therapy, spectroscopic analysis, and endoscopy, owing to its compactness, minimal invasion, large information capacity, and high resolution[1-4]. However, there are mode interference and mode coupling in MMF[5]. Therefore, target images cannot be directly transported from the proximal to the distal end of an MMF. The speckle pattern is found at the distal end face of the MMF. Subsequently, methods based on wavefront shaping[6, 7], transmission matrix(TM)[5], and phase conjugation[8] have been proposed to restore targets, but the fidelity and applicability of these methods are still limited.

As a research hotspot in recent years, the combination of deep learning(DL) and optical theory has been a great success in the fields of computational optics[9, 10]. With the advantages of data-driven neural networks, DL has excellent data fitting and mapping capabilities[11]. Intrinsic and valuable key features from speckle patterns can be extracted. Although DL shows a good way to find the mapping between speckle patterns and target information, the mapping relationship is inherent and complex, mainly representing the dataset's characteristics rather than the physical process[12]. Therefore, data-driven neural networks rely on a large-scale paired training dataset and lead to limited generalization ability[13].

Recent studies have shown that combining the physical prior of imaging systems with unsupervised training approaches has great potential to improve imaging quality[14-16]. Unsupervised training approaches have a good performance in a small-scale unpaired training dataset[17, 18]. However, from a data-driven viewpoint, there is a lack of structural similarity between the speckle pattern and the target image. Unsupervised training approaches cannot directly restore the targets[19]. Therefore, it is necessary to combine physical constraints and guidance when unsupervised networks learn and extract statistical invariants from different scattering scenes, which can effectively solve generalization problems.

In this letter, a method(TMcyclegan) based on the unsupervised network combined physical priors is proposed to restore targets at the distal end of MMF. This method has the advantage of TM and unsupervised learning. TM is a physical prior. It is used to simplify the mapping relationship between speckle patterns and target images. Then, the speckle pattern has a structural similarity with the target images, i.e., target information hidden in the speckle pattern becomes clear. At the same time, the unsupervised network(cyclegan) reconstructs target images according to the optimized direction provided by the physical prior. The research results verify the feasibility of the proposed method. Firstly, the proposed method does not need a paired training dataset. Secondly, there are only 50 in the dataset, and the unsupervised network can also applied by using only 1 speckle pattern. Finally, imaging through perturbed MMF by training only one dataset in one configuration. It's worth noting that networks trained on both MNIST and Fashion-MNIST can generalize imaging to each other. Our research is expected to advance imaging through multimode fibres from laboratory to practical applications.

As shown in Fig.1, the entire model framework consists of two components: physical priors and neural networks. After light passes through the MMF, speckle patterns can be collected by the detection device. TM can be used to characterize light propagating in MMF. So, the relationship between the speckle pattern and target information is defined as a matrix of complex coefficients $k_{mn}$. Then, m-th pixel on the speckle pattern $E_m^{speckle}$ can be expressed as[20]:

$$E_m^{speckle} = \sum_{n=1}^{N} k_{mn} E_n^{target} \quad (1)$$

where $E_n^{target}$ represents the light field at the n-th element on the target image. The linear inversion of TM optimizes the mapping relationship between the incident plane and the output plane of the fiber. The hidden target information in the speckle pattern becomes visible and clear. It can be expressed as:

$$E^{TM} = TM^{-1} E^{speckle} \quad (2)$$

where $E^{TM}$ represents the target image reconstructed by TM, $E^{speckle}$ represents the speckle pattern.

The neural networks comprise two generators and two discriminators, each dedicated to learning the bidirectional implicit relationships between high-quality reconstruction results and low-quality physical outputs. Speckle patterns captured by the CMOS utilizing TM can acquire low-quality data with similar noise distribution. Through the utilization of a bidirectional recurrent network and unpaired high-quality data, high-quality reconstruction results of hidden objects can be obtained after online training. This method does not require paired data for neural network training, allowing the reconstruction process to be completed via online training. Here, we employ a standard adversarial loss function[21] to impose constraints on both cycles, as illustrated in Eq 3, for the process of generating high-quality reconstruction result $E^{out}$ from physical data $E^{TM}$. Our target here is to optimize the generated results such that the output $G_{E^{TM} \to E^{out}}(E^{TM})$ can deceive the discriminator, making it classify them as high-quality data. In contrast, the discriminator's purpose is to optimize itself so that it can distinguish between the generated data and real high-quality data. In this process, there is an adversarial relationship between the generator and the discriminator. The same loss is also applied to the process from $E^{out}$ to $E^{TM}$, as shown in Eq 4.

$$L_{E^{TM} \to E^{out}} = \mathbb{E}(\log D_{E^{out}}(E^{out})) + \mathbb{E}(\log(1 - D_{E^{out}}(G_{E^{TM} \to E^{out}}(E^{TM})))), \quad (3)$$

$$L_{E^{out} \to E^{TM}} = \mathbb{E}(\log D_{E^{TM}}(E^{TM})) + \mathbb{E}(\log(1 - D_{E^{TM}}(G_{E^{out} \to E^{TM}}(E^{out})))), \quad (4)$$

where G is used to represent the generator in different iterations, while D is used to represent the discriminator. At the same time, we apply cycle consistency constraints $L_{cycle}$ to encourage this cyclic process.

$$L_{cycle} = \mathbb{E}\left(\left\|G_{E^{out} \to E^{TM}}\left(G_{E^{TM} \to E^{out}}(E^{TM})\right) - E^{TM}\right\|_1\right) + \mathbb{E}\left(\left\|G_{E^{TM} \to E^{out}}\left(G_{E^{out} \to E^{TM}}(E^{out})\right) - E^{out}\right\|_1\right), \quad (5)$$

Therefore, the total loss function L is as follows:

$$L = L_{E^{TM} \to E^{out}} + L_{E^{out} \to E^{TM}} + L_{cycle}. \quad (6)$$

To emphasize the generalizability of this approach to image through MMF under unknown configurations, unpaired data is used during the training process, making the entire procedure unsupervised. This implies that the trained model can be applied to imaging through unknown configurations while also highlighting the advantages of physics-driven approaches. All the neural networks mentioned in this paper are running on the computing platform with GPU RTX3090Ti. The network is run on Ubuntu16.04 and is accelerated by Pytorch1.8 and CUDA11.1.

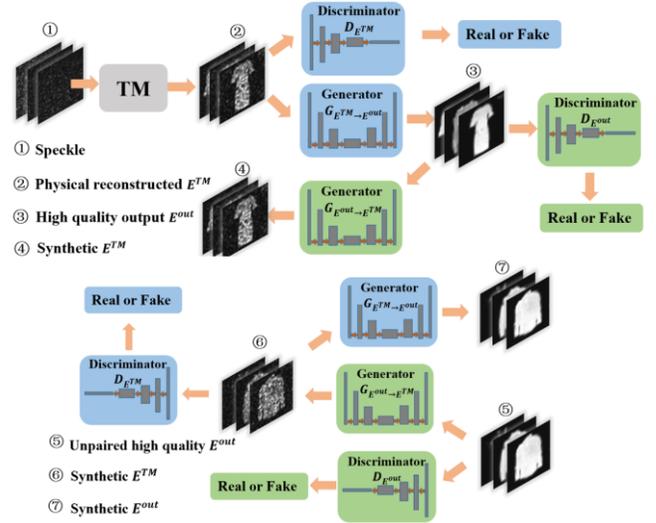

Fig.2. Unsupervised framework structure diagram for TM-driven imaging through scattering media

To verify the effectiveness of the proposed method, the system shown in Fig. 2 is used to collect speckle patterns. The light is emitted by the 532 nm continuous laser (LCX-532S, Oxxius, 80 mW). The beam expander (GBE20-A-20×, Thorlabs, 20x) expands and collates light. Polarizer P is used to select the appropriate polarization state for the spatial light modulator (SLM, PLUTO-2-vis-096, HOLOEYE, pixel pitch: 8 μm). Then, SLM is used to code and display the 8-bit objects. The modulated light is scaled by the lens((L1: AC254-200, Thorlabs, f:200mm; L2: AC254-150, Thorlabs, f: 150mm). The light can be coupled into an MMF (DH-FMM200-FC-1AF, Daheng Optics, NA:0.22, Core Diameter:200um) by the objective (RMS10X, Thorlabs, NA:0.25). The scattered light from the MMF is collected by the other objective (RMS20X, Thorlabs, NA:0.4) and finally detected by a CMOS (BFS-U3–04S2M-CS, Point Gray; pixel pitch: 6.9 μm).

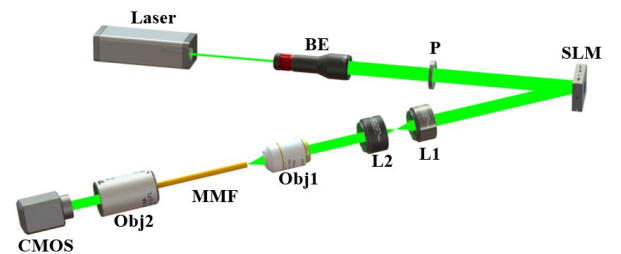

Fig.2. Schematic diagram of the experimental setup. BE: beam expander; L1, L2: lens; P: polarizer; OBJ1 and OBJ2: objective lens; SLM: spatial light modulator; MMF: multimode fiber.

The target images are picked from the modified MNIST[22] and Fashion-MNIST datasets[23]. Before the experiment, we fill in the "0" pixels to expand the original target images to 32×32 pixel images. The data can be roughly divided into 6

groups($D\_1$-$D\_6$) according to different configurations and datasets. In each group, 50 target images were used to generate speckle patterns according to the experimental equipment in Fig.2. The speckle patterns captured by the CMOS from the distal end of the fiber need to be cropped into 400*400 pixels and then downsampled to $32 \times 32$ pixels as training samples. Then, speckle patterns and the other 50 target images comprise the unpaired training dataset. Specific details are shown in table 1. In order to evaluate the superiority of the proposed method, two models based on the Unet are trained[24], named Unet and TM-Unet, respectively.

**Table 1. Details of different training sets.**

| Type | Dataset | Size |
|---|---|---|
| MNIST | $D\_1,D\_2,D\_3$ | 50 |
| Fashion-MNIST | $D\_4,D\_5,D\_6$ | |

Fig.3. Comparison of reconstruction results between the Unet, TM-Unet, and the proposed unsupervised method under one configuration.

The reconstruction ability of the proposed method is investigated. $D\_1$ and $D\_4$ are used to train the TMcyclegan, Unet, and TM-unt, respectively. Their reconstruction results are shown in Fig. 3. Unet cannot reconstruct the target image at all. Target information in the speckle patterns cannot be extracted because of a small-scale training dataset. The mapping relationship between speckle patterns and target images cannot be created. The reconstruction results of TM-Unet are a little better than those of Unet, but their structures remain indistinguishable. Although TM simplifies the mapping relationship, the optimization direction of TM-Unet is confounded by the unpaired training dataset. TMcyclegan does not rely on a large-scale paired training dataset. Its reconstruction ability is far better than the other two. PCC, SSIM, and PSNR were used to quantitatively evaluate the reconstructed images' quality. PCC, SSIM, and PSNR of the results are 0.986， 0.94, and 23.9dB, respectively, when training MNIST. PCC, SSIM, and PSNR of the results are 0.973， 0.914, and 21.3dB, respectively, when training Fashion-MNIST. The structure of the target image is well reconstructed. The results of 6 training sets are shown in Fig.S1 of the supplementary material. The target image can be reconstructed well. Also, in Fig.S2 of the supplementary material, the reconstruction results of Cyclegan can be found. Cyclegan cannot reconstruct the target image directly from the speckle pattern.

Fig.4. Comparison of generalization capability between the Unet, TM-Unet, and the proposed unsupervised method.

To verify the generalization capability of the proposed method, $D\_1$ and $D\_4$ are used to train networks. Other groups are used to test. Mapping relationships between the speckle pattern and the target image are different when MMF is under different configurations. The target image cannot be reconstructed when using a mismatched mapping relation, as shown in Fig.4. Neither Unet nor TM-Unet was able to reconstruct the target images in the test set. Some researchers have made some progress by training with target-speckle pairs in several configurations. However, networks need more data than before. TMcyclegan has a strong generalization ability. It only needs a training set under a configuration to establish a generic mapping relationship. When training with $D\_1$ or $D\_4$, the target images in $D\_2$, $D\_3$, $D\_5$, and $D\_6$ can be reconstructed by the TMcyclegan. Quality evaluation of the reconstruction results is shown in Table 2. TMcyclegan can also image across different types of data. The data type of $D\_1$ is the same as that of $D\_2$ and $D\_3$. The data type of $D\_4$ is the same as that of $D\_5$ and $D\_6$. When using $D\_1$ to train TMcyclegan, the target images in $D\_2$ and $D\_3$ can be reconstructed well. Their PCC, SSIM, and PSNR were much higher than the reconstruction of $D\_5$ and $D\_6$. Because the data type of $D\_2$ and $D\_3$ is the same as that of $D\_1$, but that of $D\_5$ and $D\_6$ are not. When using $D\_4$ to train TMcyclegan,

PCC, SSIM, and PSNR of D_5 and D_6 increased significantly. However, the PCC, SSIM, and PSNR of D_2 and D_3 don't drop much. The data types of Fashion-MNIST are more complex than those of MNIST. Thus, TMcyclegan has learned the complex logic of composition. Then, the mapping between speckle patterns and target information becomes effective and complete.

**Table 2. PCC, SSIM, and PSNR of the reconstruction results.**

| Training Set | Testing Set | PCC | SSIM | PSNR(dB) |
|---|---|---|---|---|
| D_1 | D_2 | 0.972 | 0.936 | 23.6 |
| | D_3 | 0.983 | 0.951 | 22.9 |
| | D_5 | 0.901 | 0.778 | 15.3 |
| | D_6 | 0.854 | 0.730 | 16.4 |
| D_4 | D_2 | 0.963 | 0.898 | 20.0 |
| | D_3 | 0.980 | 0.925 | 21.7 |
| | D_5 | 0.966 | 0.861 | 22.6 |
| | D_6 | 0.957 | 0.879 | 22.6 |

In this letter, a physics-assisted, unsupervised, learning-based fibre imaging scheme named TMcyclegan is proposed and demonstrated, which combines the advantages of physical prior and unpaired-data-driven approaches. TM is used to decode the target information in the speckle pattern in advance, which simplifies the mapping between the speckle pattern and the target image. Then, cyclegan is used to learn the propagation process of the target in the MMF by a small-scale unpaired training dataset. Experiments present the outstanding capabilities of the TMcyclegan. Firstly, TMcyclegan does not require a large-scale paired dataset, which is a critical factor in many practical applications. Secondly, it exhibits strong robustness to against the influence of perturbed MMF. Finally, it easily generalized to high-fidelity imaging when the network trains different datasets. However, when the type of training dataset is simple, TM's ability to reconstruct complex targets will be reduced. This method may not be able to reconstruct the details of complex targets effectively. We hope that the combination of physical prior and unsupervised networks can accelerate the transition of MMF imaging from laboratory research to practical applications.

**Funding.** National Nature Science Foundation of China (No.62362037）, Natural Science Foundation of Jiangxi Province of China (No.20224ACB202011).

**Acknowledgments.**

**Disclosures.** The authors declare no conflicts of interest.

**Data availability.** Data underlying the results presented in this paper are not publicly available at this time but may be obtained from the authors upon reasonable request.

**References**
1. Y. Choi, C. Yoon, M. Kim et al., Phys Rev Lett **109**, 203901 (2012).
2. A. M. Caravaca-Aguirre, and R. Piestun, Opt Express **25**, 1656-1665 (2017).
3. S. Turtaev, I. T. Leite, T. Altwegg-Boussac et al., Light: Science & Applications **7** (2018).
4. C. Zhang, Z. Yao, T. Liu et al., Optics & Laser Technology **169** (2024).
5. M. Plöschner, T. Tyc, and T. Čižmár, Nature Photonics **9**, 529-535 (2015).
6. O. Tzang, A. M. Caravaca-Aguirre, K. Wagner et al., Nature Photonics **12**, 368-374 (2018).
7. C. Zhang, Z. Yao, Z. Qin et al., Optics and Lasers in Engineering **164** (2023).
8. I. N. Papadopoulos, S. Farahi, C. Moser et al., Optics Express **20**, 10583-10590 (2012).
9. G. Barbastathis, A. Ozcan, and G. Situ, Optica **6** (2019).
10. Y. LeCun, Y. Bengio, and G. Hinton, Nature **521**, 436-444 (2015).
11. S. Li, M. Deng, J. Lee et al., Optica **5** (2018).
12. S. Zhu, E. Guo, J. Gu et al., Photonics Research **9** (2021).
13. S. Gigan, O. Katz, H. B. de Aguiar et al., Journal of Physics: Photonics **4** (2022).
14. Y. Shi, E. Guo, M. Sun et al., Optics Letters **47** (2022).
15. S. Liu, X. Meng, Y. Yin et al., Optics and Lasers in Engineering **147** (2021).
16. Y. Shi, E. Guo, M. Sun et al., Results in Physics **51** (2023).
17. K. Yamazaki, R. Horisaki, and J. Tanida, Applied Optics **59** (2020).
18. X. Hu, J. Zhao, J. E. Antonio-Lopez et al., Optics Express **31** (2023).
19. Y. Shi, E. Guo, L. Bai et al., Optics Express **30** (2022).
20. S. M. Popoff, G. Lerosey, R. Carminati et al., Phys Rev Lett **104**, 100601 (2010).
21. W. Li, K. Abrashitova, G. Osnabrugge et al., Physical Review Applied **18** (2022).
22. Y. LeCun；, C. Cortes；, and C. J. C. Burges, "The mnist database of handwritten digits."
23. R. K. Xiao H, Vollgraf R, (2017).
24. H. Gao, H. Hu, Y. Zhang et al., Optics & Laser Technology **167** (2023).